\documentclass[conference]{IEEEtran}
\IEEEoverridecommandlockouts
\usepackage{cite}
\usepackage{amsmath,amssymb,amsfonts}
\usepackage{algorithm}
\usepackage{algpseudocode}
\usepackage{graphicx}
\usepackage{float}
\usepackage{textcomp}
\usepackage{xcolor}
\def\BibTeX{{\rm B\kern-.05em{\sc i\kern-.025em b}\kern-.08em
    T\kern-.1667em\lower.7ex\hbox{E}\kern-.125emX}}
\begin{document}


\title{Adversarially-Informed Node Criticality Identification in Power Grid Measurements\\
}

\author{\IEEEauthorblockN{1\textsuperscript{st} Koto Omiloli}
\IEEEauthorblockA{\textit{Department of Electrical  Engineering} \\
\textit{Florida State University}\\
Tallahassee, FL 32310, USA \\
 Email: kao23a@fsu.edu}
\and
\IEEEauthorblockN{2\textsuperscript{nd} Olugbenga Moses Anubi}
\IEEEauthorblockA{\textit{Department of Electrical  Engineering} \\
\textit{Florida State University}\\
Tallahassee, FL 32310, USA \\
Email: oanubi@fsu.edu}
}
\newtheorem{definition}{Definition}
\maketitle
\begin{abstract}
Power grid state estimation relies on sensor measurements that are increasingly vulnerable to adversarial corruption in cyber--physical environments, potentially leading to significant deviations in system observations. This motivates the need to identify critical measurement nodes whose compromise results in the most severe system-level impact. However, existing node criticality methods primarily rely on structural or steady-state analyses and do not explicitly account for adversarial effects on system behavior. To address this gap, this paper proposes an adversarially informed framework for identifying critical measurement nodes in linearized power systems. Within this framework, a structured attack generation mechanism is developed to construct stealthy and effective false data injection attacks (FDIAs) against an $H_\infty$ resilient state estimator. Node criticality is then evaluated using coalition-based marginal contributions of compromised sensor subsets, estimated via permutation sampling over a prescribed set of admissible nodes, with the resulting importance scores mapped to the corresponding physical buses. Simulation results on the IEEE 14-bus system show that adversarially identified nodes induce larger deviations in frequency, voltage angle, and net power compared to randomly selected nodes, demonstrating the effectiveness of the proposed framework.
\end{abstract}


\section{Introduction}
Modern power systems are increasingly operated as cyber–physical infrastructures in which sensing, communication, and control layers are tightly integrated with physical grid dynamics. While this integration improves observability and operational efficiency, it also introduces new vulnerabilities, particularly through adversarial manipulation of measurement data used in state estimation and control~\cite{omiloli2026}. Even localized corruption of sensor measurements can propagate through system dynamics and degrade estimation accuracy, potentially leading to incorrect control actions and large-scale instability~\cite{9496085}. This motivates the need for systematic identification of critical measurement nodes~\cite{megzari2023applications} whose compromise has disproportionate impact on grid observability and stability.

The notion of node criticality in power systems has been extensively studied in the context of structural vulnerability~\cite{ugurlu2025critical}, dynamic performance, and interdependent infrastructure resilience. Early work on interdependent networks highlights how cascading failures between power and communication layers significantly amplify system fragility, framing critical node identification as a coupled optimization problem over interacting networks~\cite{nguyen2013detecting}. Subsequent approaches have incorporated steady-state and dynamic operational criteria, combining optimal power flow and transient stability analysis to rank critical buses under uncertainty~\cite{7094325}. These methods demonstrate that node importance is strongly dependent on operating conditions, particularly load variability and generator dispatch.

More recent approaches incorporate electrical and graph-theoretic structure into node ranking. Electrical topology and power flow sensitivity have been integrated using centrality-based formulations, including PageRank-inspired models that account for power transfer dependencies~\cite{fan2022critical}. In parallel, control-theoretic perspectives have been introduced through structured network representations such as the electric cactus framework, which defines controllability- and observability-based indices for ranking nodes according to their influence on system dynamics~\cite{8981873}. These developments extend classical graph-based measures by incorporating physically meaningful electrical interactions.

Cyber–physical extensions of these models further emphasize the coupling between communication and power layers. In particular, multi-perspective evaluation frameworks for cyber–physical power systems demonstrate that node importance varies depending on whether structural, informational, or electrical characteristics are considered~\cite{li2024methods}. However, such methods remain primarily descriptive and do not explicitly model adversarial manipulation of measurement data or its effect on state estimation performance.

In parallel, adversarial and control-oriented studies have examined attack strategies targeting dynamical systems. Multi-objective formulations have been used to characterize attacker behavior in terms of energy, sparsity, and convergence time~\cite{faramondi2019assessing}, while cascading failure models based on self-organized criticality capture nonlinear propagation effects in power grids~\cite{mei2008study}. Nevertheless, these frameworks generally focus on system-level disruption or structural failure propagation and do not explicitly connect adversarial sensor attacks to node-level criticality in state estimation.

Despite these advances, three key limitations remain in existing literature. First, most approaches evaluate node importance using either structural topology, steady-state power flow sensitivity, or predefined dynamic scenarios, but rarely integrate these perspectives within a unified framework. Second, adversarial effects are typically modeled at the system level or as exogenous disturbances, without explicitly linking measurement corruption to node-level criticality. Third, existing frameworks do not explicitly quantify how sensor compromise affects state estimation quality, despite the central role of estimation in modern grid monitoring and control.

To address these limitations, this paper proposes an adversarially informed framework for critical node identification in power grids. Node importance is assessed by analyzing how stealthy measurement attacks influence system behavior against an $H_\infty$ state estimation framework. This is enabled by a structured attack generation scheme that produces bounded perturbations, and a coalition-based marginal contribution metric that ranks sensor nodes by impact.

The main contributions of this work are summarized as follows:
\begin{itemize}
    \item A unified adversarial framework for measurement node criticality assessment that integrates linearized power system dynamics, resilient state estimation, and adversarial measurement corruption.
    \item A coalition-based vulnerability metric that quantifies node importance through marginal contributions of compromised sensor subsets under adversarial conditions.
    \item Comprehensive simulation validation on the IEEE 14-bus system demonstrating that adversarially ranked critical nodes induce substantially greater degradation in grid dynamic performance than randomly selected nodes.
\end{itemize}

The remainder of this paper is organized as follows. Section II introduces the notations. Section III presents the linearized power grid model. Section IV develops the $H_\infty$ resilient state estimation framework. Section V describes the structured attack generation methodology. Section VI formulates the node criticality assessment approach. Section VII provides simulation results on the IEEE 14-bus system. Finally, Section VIII concludes the paper and discusses future research directions.



\section{Notation}
We denote $\mathbb{R}$ and $\mathbb{R}^n$ as the set of real numbers and the space of real-valued vectors of dimension $n$, respectively. Scalars are denoted by lowercase letters (e.g., $x \in \mathbb{R}$), vectors by bold lowercase letters (e.g., $\mathbf{x} \in \mathbb{R}^n$) with entries $x_i$, and matrices by uppercase letters (e.g., $K \in \mathbb{R}^{n \times m}$). The zero vector is denoted by $\mathbf{0}$, the all-ones vector by $\mathbf{1}$, and the identity matrix by $I$. Let $\mathcal{V} = \{1,2,\dots,n\}$ index the grid sensor nodes; for $S \subseteq \mathcal{V}$, $A_S$ denotes the matrix $A$ with rows outside $S$ set to zero. The notation $\mathcal{N}(\mathbf{0}, I)$ denotes the Gaussian distribution with zero mean and identity covariance matrix.

\begin{definition}[LeakyReLU]
The Leaky Rectified Linear Unit (LeakyReLU) activation function is defined as
$
\mathrm{LeakyReLU}(x) =
\begin{cases}
x, & x \ge 0, \\
\alpha x, & x < 0,
\end{cases}
$
where $0 < \alpha \ll 1$.
\end{definition}

\section{Grid Network Modelling}
The model considered in this work is a linearized networked dynamical system obtained by linearizing the generator swing equations and power flow equations around a steady-state operating point under standard small-signal and DC power flow assumptions. The resulting state-space representation, adapted from~\cite{anubi2020multi} and extended to include dynamic load models, is given by:
\begin{equation}\label{grid_model}
\begin{split}
\dot{\mathbf{x}} &= 
\underbrace{
\begin{bmatrix}
0 & I & 0 \\
- M^{-1}L_{gg} & -M^{-1}D_g & -M^{-1}L_{gl} \\
- D_l^{-1}L_{lg} & 0 & -D_l^{-1}L_{ll}
\end{bmatrix}
}_{A}
\mathbf{x} \\
&\quad +
\underbrace{
\begin{bmatrix}
0 & 0 \\
M^{-1} & 0 \\
0 & D_l^{-1}
\end{bmatrix}
}_{B}
\mathbf{u}, \\[6pt]
\mathbf{y} &= 
\underbrace{
\begin{bmatrix}
I & 0 & 0 \\
0 & I & 0 \\
L_{gg} & 0 & L_{gl} \\
L_{lg} & 0 & L_{ll}
\end{bmatrix}
}_{C}
\mathbf{x} + \mathbf{y}^a,
\end{split}
\end{equation}
where
$
\mathbf{x} =
\begin{bmatrix}
\boldsymbol{\theta}_g^\top & \boldsymbol{\omega}_g^\top & \boldsymbol{\theta}_l^\top
\end{bmatrix}^\top \in \mathbb{R}^{2n_g + n_l}
$
consists of generator rotor angles \(\boldsymbol{\theta}_g \in \mathbb{R}^{n_g}\), frequency deviation \(\boldsymbol{\omega}_g \in \mathbb{R}^{n_g}\), and load bus voltage angles \(\boldsymbol{\theta}_l \in \mathbb{R}^{n_l}\). The input vector
$
\mathbf{u} =
\begin{bmatrix}
\mathbf{p}_m^\top & \mathbf{p}_d^\top
\end{bmatrix}^\top
$
represents mechanical power injections \(\mathbf{p}_m \in \mathbb{R}^{n_g}\) and active loads \(\mathbf{p}_d \in \mathbb{R}^{n_l}\). The output
$
\mathbf{y} =
\begin{bmatrix}
\boldsymbol{\theta}_g^\top &   \boldsymbol{\omega}_g^\top & \mathbf{p}_{net}^\top
\end{bmatrix}^\top
$
consists of generator frequency measurements and net power injections \(\mathbf{p}_{net} \in \mathbb{R}^{n_b}\), corrupted by measurement attacks \(\mathbf{y}^a\). The matrices $M \in \mathbb{R}^{n_g \times n_g}$, $D_g \in \mathbb{R}^{n_g \times n_g}$, and $D_l \in \mathbb{R}^{n_l \times n_l}$ denote the generator inertia, generator damping, and load damping matrices, respectively, defined as
$
M = \mathrm{diag}(m_1, \dots, m_{n_g}), \quad
D_g = \mathrm{diag}(d^g_1, \dots, d^g_{n_g}), \quad
D_l = \mathrm{diag}(d^l_1, \dots, d^l_{n_l}),
$
where $m_i$, $d_i^g$  denotes the inertia constant, damping coefficient of the $i$-th generator respectively and $d_i^l$ denotes the damping coefficient of the $i$-th load bus. The network is represented by the Laplacian matrix $L = \mathrm{diag}(B\mathbf{1}) - B$, where $B$ is the network susceptance matrix. The buses are ordered such that generator buses precede load buses, yielding the partitioned structure
$
L =
\begin{bmatrix}
L_{gg} & L_{gl} \\
L_{lg} & L_{ll}
\end{bmatrix}.
$

\section{Resilient Network State Estimation}

An $H_\infty$ resilient observer for the grid network model in \eqref{grid_model} is developed to ensure reliable state estimation for secure and stable grid operation under adversarial measurement corruption. To this end, the underlying continuous-time dynamics are first discretized using a sampling period $T_s$, yielding the corresponding discrete-time representation:
\begin{equation}
\begin{aligned}
\mathbf{x}_{k+1} &= \tilde{A}\mathbf{x}_k + \tilde{B}\mathbf{u}_k, \\
\mathbf{y}_k &= C\mathbf{x}_k + \mathbf{y}_k^a,
\end{aligned}
\end{equation}
where $\tilde{A} = I + A T_s$ and $\tilde{B} = B T_s$.

To estimate the grid states, we consider a Luenberger-type observer of the form:
\begin{equation}
\hat{\mathbf{x}}_{k+1} =
\tilde{A}\hat{\mathbf{x}}_k + \tilde{B}\mathbf{u}_k
+ L(\mathbf{y}_k - C \hat{\mathbf{x}}_k),
\label{xhat}
\end{equation}
where $L$ is the observer gain to be designed.

Hence, defining the estimation error $\mathbf{e}_k = \mathbf{x}_k - \hat{\mathbf{x}}_k$, its dynamics evolve as:
\begin{equation}
\mathbf{e}_{k+1} =
(\tilde{A} - LC)\mathbf{e}_k - L\mathbf{y}_k^a.
\end{equation}

The objective is to design  $L$ such that the effect of the attack input $\mathbf{y}_k^a$ on the estimation error is attenuated in the $H_\infty$ sense, that is:
\begin{equation}
\|\mathbf{e}_k\|_2 \le \gamma \|\mathbf{y}_k^a\|_2, \quad \gamma > 0,
\end{equation}
ensuring robustness against worst-case measurement attacks~\cite{wang1992observer}.

Now consider the quadratic Lyapunov candidate function:
\begin{equation}
V_k = \mathbf{e}_k^\top P \mathbf{e}_k, \quad P = P^\top \succ 0.
\end{equation}

It is well known that the $H_\infty$ performance requirement is satisfied if the following dissipation inequality~\cite{10644849} holds:
\begin{equation}
V_{k+1} - V_k + \mathbf{e}_k^\top \mathbf{e}_k
- \gamma^2 (\mathbf{y}_k^a)^\top \mathbf{y}_k^a < 0.
\label{dissipation}
\end{equation}

Substituting the error dynamics into \eqref{dissipation} and rearranging terms yields a quadratic form in $(\mathbf{e}_k, \mathbf{y}_k^a)$:
\begin{align}
\begin{bmatrix}
\mathbf{e}_k \\
\mathbf{y}_k^a
\end{bmatrix}^\top
\mathcal{M}
\begin{bmatrix}
\mathbf{e}_k \\
\mathbf{y}_k^a
\end{bmatrix}
< 0,\label{error_condition}
\end{align}
where
\begin{align}
\mathcal{M} =
\begin{bmatrix}
M_{11} & M_{12} \\
M_{12}^\top & M_{22}
\end{bmatrix},
\label{M_matrix}
\end{align}

with the block matrices given by
$M_{11}  = (\tilde{A}-LC)^\top P (\tilde{A}-LC) - P + I, 
M_{12}  = -(\tilde{A}-LC)^\top P L, \text{and }
M_{22} = L^\top P L - \gamma^2 I. $

Factorizing the quadratic terms and introducing the change of variables $Q = PL$, the matrix in \eqref{M_matrix} is expressed as:
\begin{align}\mathcal{M} =
    \begin{bmatrix}
        H_{11} &  H_{12} \\
        H_{12}^\top &  H_{22}
    \end{bmatrix} + 
    \begin{bmatrix}
    QC & Q
\end{bmatrix}^\top P^{-1}
\begin{bmatrix}
    QC & Q
\end{bmatrix},
\end{align}
where
$H_{11} = \tilde{A}^\top P \tilde{A} - C^\top Q^\top \tilde{A} - \tilde{A}^\top Q C - P + I,\quad
H_{12} = -\tilde{A}^\top Q,\quad
H_{22} = -\gamma^2 I.$

Finally, using the Schur compliment, we construct a matrix $\bar{\mathcal{M}}$ such that
$\bar{\mathcal{M}} \prec 0$ is equivalent to $\mathcal{M} \prec 0$: 
\begin{equation}
\bar{\mathcal{M}}(\gamma, P, Q) =
\begin{bmatrix}
H_{11} & -\tilde{A}^\top Q & C^\top Q^\top \\
-Q^\top \tilde{A} & -\gamma^2 I & Q^\top \\
Q C & Q & -P
\end{bmatrix}.
\end{equation}
The observer design is then formulated as the following optimization problem:
\begin{equation}
\begin{aligned}
\min_{\gamma, P, Q} \quad & \gamma^2 \\
\text{s.t.} \quad & \bar{\mathcal{M}}(\gamma, P, Q) \prec 0, \\
& P \succ 0,
\end{aligned}
\end{equation}
from which the observer gain is recovered as $L = P^{-1}Q$.

\section{Attack Generation Modelling} 
We consider structured attack generation for assessing grid sensor node criticality by adopting our prior framework in \cite{zheng2026generative, zheng2023data}, where FDIAs were modeled as sparse attacks injected into system measurements. The set of feasible attacks was defined as:
\begin{equation}\label{constraints}
 \mathcal{S} = \left\{ \mathbf{y}_k^a \in \Sigma_k \mid e(\mathbf{y}_k^a) \geq \tau_e,\; s(\mathbf{y}_k^a) \leq \tau_s \right\},
\end{equation}
where $\Sigma_k$ denotes the set of $k$-sparse attack vectors, and
\begin{equation}
    e(\mathbf{y}_k^a) = \|\mathbf{y}_k\|, \quad
s(\mathbf{y}_k^a) = \|\hat{\mathbf{x}}_k^a - \hat{\mathbf{x}}_k\|,
\end{equation}

represent the effectiveness and stealthiness metrics, respectively, with thresholds $\tau_e$ and $\tau_s$.


where $\alpha \in (0,1)$, $\theta_G$ denotes the generator weights.

In this work, the framework is applied to node criticality assessment by restricting attack injection to admissible measurement nodes and quantifying the resulting degradation in state estimation performance under the effectiveness and stealth criteria. The attack generation framework comprises six components: a generator, two discriminators (effectiveness and stealthiness), the physics-based grid model, an attack policy, and resilient $H_\infty$ estimator. During training, real-time system measurements are used to guide the discriminator networks, which in turn supervise the generator to produce feasible attacks.

The generator is trained by minimizing the loss
\begin{equation}
\mathcal{L}_G =
\log\!\left(
1 + \max\!\left(
0,
\mathbb{E}_{\mathbf{z}\sim \mathcal{N}}
\big[D(G(\mathbf{z};\theta_G))\big] - (1-\alpha)
\right)\right),
\label{generator_loss}
\end{equation}
and \(D(\cdot)\) is defined as:
\begin{align}
D(\mathbf{y}_k^a) &=
\exp \Big(
\textsf{LeakyReLU}(\tau_e - f_1(\mathbf{y}_k^a; \theta_1)) \nonumber \\
&\quad +
\textsf{LeakyReLU}(f_2(\mathbf{y}_k^a; \theta_2) - \tau_s)
\Big),
\end{align}
with $f_1$ and $f_2$ representing the effectiveness and stealthiness discriminator networks, respectively, parameterized by their corresponding weights $\theta_1$ and $\theta_2$.

Hence, the generator is trained using the sample-based optimization:
\begin{equation}
\theta_G^* =
\arg\min_{\theta_G}
\mathbb{E}_{\mathbf{z}\sim P_{\mathbf{z}}}
\big[\mathcal{L}_G(G(\mathbf{z};\theta_G))\big].
\end{equation}

Similarly, the discriminators are trained to approximate the grid model’s effectiveness and stealthiness outputs using a mean squared error loss.
\begin{equation}
\theta_i^* =
\arg\min_{\theta_i}
\frac{1}{m} \sum_{k=1}^{m}
\big\|
l_i(\mathbf{y}_k^a) - f_i(\mathbf{y}_k^a;\theta_i)
\big\|^2,
\quad i \in \{1,2\}.
\label{mean_squared_error}
\end{equation}

The attacks are then generated according to the following policy:
\begin{align}
\pi_r(t;\mathbf{i}) &=
\begin{cases}
0, & t < t_0, \\
\min\!\left(\frac{t - t_0}{d}, 1\right) a, & \text{otherwise},
\end{cases}
\label{attack_policy}
\end{align}
where $\mathbf{i} \in \mathbb{R}^4$ is defined as the attack vector for an individual measurement node, extracted from the generator output $\mathbf{i}_g \in \mathbb{R}^{4n}$, with $n$ denoting the number of measurement nodes attacked. The policy parameters are thus defined as:
\begin{align}
t_0 &= t_s + (t_f - t_s)i_1, \quad
d = d_s + (d_f - d_s)i_2, \nonumber \\
t_1 &= t_{1_s} + (t_{1_f} - t_{1_s})i_3, \quad
a = a_s + (a_f - a_s)i_4,
\end{align}

where \(t_0\) denotes the attack start time, \(d\) the duration, \(t_1\) the time span, and \(a\) the attack magnitude, with the attack policy parameters \(\theta_p = (t_s, t_f, d_s, d_f, t_{1_s}, t_{1_f}, a_s, a_f)\) defining admissible ranges.  The resulting per-node attack signals populate the grid measurement attack vector \(\mathbf{y}_k^a \in \mathbb{R}^{2n_g+n_b}\) over an attack support. For $s_g$ number of generator samples, the attack data is built as $A = [\mathbf{t}_0\quad \mathbf{t}_{t1}\quad \mathbf{d} \quad \mathbf{a}] = \mathcal{T}(I_g, \theta)$ where $ A, I_g \in \mathbb{R}^{4n \times s_g}$. 


\section{Node Criticality Assessment}

\begin{algorithm}[htpb]
\caption{Node Criticality Assessment}
\label{alg:node_criticality}
\begin{algorithmic}[1]

\Procedure{$f$}{$G, D_s, D_e, s_g, \theta_p, \tau_s, n, M$}

\State Sample $Z \sim \mathcal{N}(\mathbf{0},I)$
\State $I_g \leftarrow G(Z)\;$ \Comment{generator output}
\State $A \leftarrow \mathcal{T}(I_g, \theta_p)$  \Comment{construct attack data}
\State Generate $M$ permutations $\{\pi_m\}$ over $\mathcal{V}$

\State Initialize $\phi \leftarrow \mathbf{0} \in \mathbb{R}^{n}$

\For{each permutation $\pi_m$}
    \State $S \leftarrow \emptyset$, $v(S) \leftarrow 0$
    
    \For{each $j \in \pi_m$}
        \State $S' \leftarrow S \cup \{j\}$ \Comment{current coalition}
        \State $\mathbf{e} \leftarrow D_e(A_{S'})$, $\mathbf{s} \leftarrow D_s(A_{S'})$ \Comment{$\in \mathbb{R}^{s_g}$}
        
        \State $\mathcal{I} \leftarrow \{k \in \{1,\dots,s_g\} : s_k \le \tau_s\}$ \Comment{feasible samples}
        
        \If{$\mathcal{I} \neq \emptyset$}
            \State $v(S') \leftarrow \max\limits_{k \in \mathcal{I}}  e_k$
        \Else
            \State $v(S') \leftarrow \max\limits_{k} e_k$
        \EndIf
        
        \State $\phi_j \leftarrow \phi_j + \big(v(S') - v(S)\big)$
        \State $S \leftarrow S'$
    \EndFor
\EndFor

\State $\phi \leftarrow \phi / M$  \Comment{average over sampled permutations}

\State \Return $\phi$ 

\EndProcedure

\end{algorithmic}
\end{algorithm}

To quantify grid node criticality, permutations of the admissible nodes are sampled and coalitions are formed incrementally by adding one node at a time. For each coalition, the marginal contribution of node $j$ is computed as $v(S)-v(S\setminus\{j\})$, where $v(S)$ measures the vulnerability induced by compromising the sensor subset $S\subseteq\mathcal{V}$. Specifically, the coalition vulnerability metric is defined as:
\[
v(S)=\max_{k:\,s_k\le\tau_s} e_k,\quad k\in\{1,\dots,s_g\},
\]

where $k$ indexes the generated attack samples, and $e_k$ and $s_k$ denote the impact and stealthiness of the $k$th sample, respectively. The criticality score of node $j$, denoted by $\phi_j$, is then obtained by averaging its marginal contributions across all sampled permutations, yielding a quantitative ranking of sensor importance that are mapped to their corresponding buses for system-level vulnerability assessment.

The proposed framework is summarized in Algorithm~\ref{alg:node_criticality}, which takes as input the trained generator network $G$, the trained discriminators $D_s$ and $D_e$ for stealthiness and effectiveness evaluation, respectively, the number of generator samples $s_g$, the attack policy parameters $\theta_p$, the stealthiness threshold $\tau_s$, the number of admissible nodes $n$, and the number of sampled permutations $M$. 





    
        
        





\section{Simulation Results}
\begin{figure}
    \centering
    \includegraphics[scale=0.3]{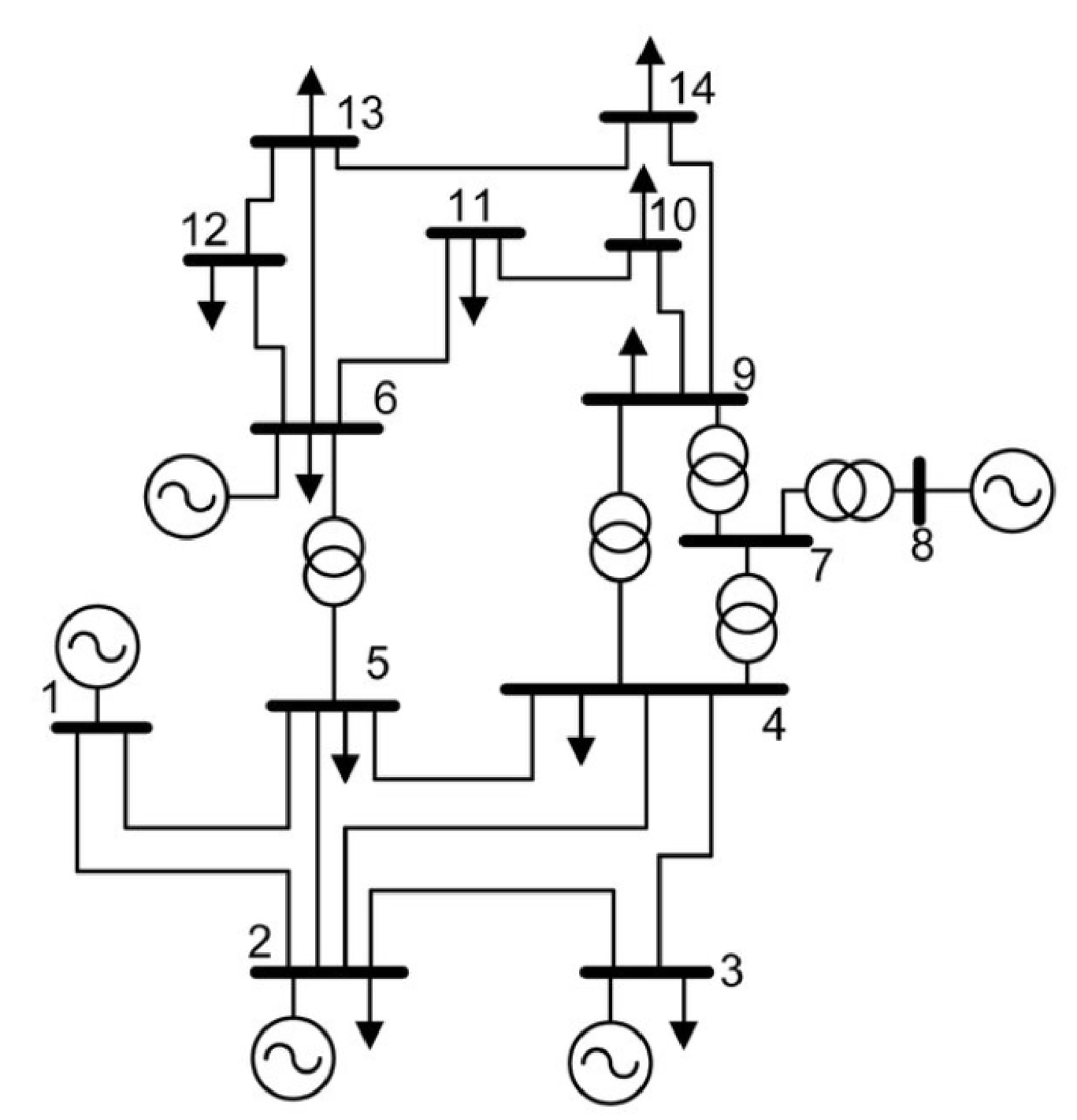}
    \caption{IEEE 14-bus power system test network single-line diagram~\cite{tortos2013steady}.}
    \label{IEEE_14bus}
\end{figure}

The proposed node criticality framework was evaluated using a linearized dynamic model of the IEEE 14-bus power system shown in Fig. \ref{IEEE_14bus}. Network parameters were derived from the IEEE 14-bus test case to construct the system Laplacian and associated state-space representation. The generator inertia and damping matrices were set as $M=\mathrm{diag}(2.1,2.2,2.3,2.4,2.5)$ and $D_g=\mathrm{diag}(0.72,0.71,0.73,0.65,0.70)$, respectively, and the load damping matrix was defined as $D_l=\mathrm{diag}(0.22,0.21,0.33,0.25,0.40)$. Active loads $\mathbf{p}_d$ were fixed to nominal demand values, and mechanical inputs $\mathbf{p}_m$ were designed via output feedback to regulate frequency deviations to zero. The $H_\infty$ resilient observer was synthesized using YALMIP~\cite{lofberg2004yalmip} with SDPT3~\cite{toh1999sdpt3} as the solver. 

\begin{figure}[htpb]
    \centering
    \includegraphics[scale=0.7]{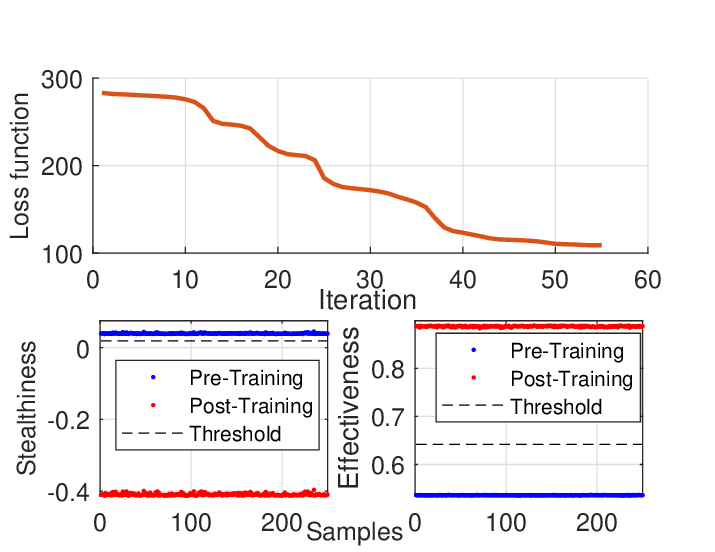}
    \caption{Generator loss (top); stealthiness (bottom left) and effectiveness (bottom right) evaluated against thresholds on 25\% of generated samples.}
    \label{trainingloss}
\end{figure}

For the attack generation, parameters were set as $n=24$, $s_g=1000$, $(\tau_e, \tau_s) = (0.64, 0.04)$, with attack policy ranges $(t_s,t_f)=(3,9)$, $(t_{1_s},t_{1_f})=(2,6)$, $(d_s,d_f)=(3,20)$, and $(a_s,a_f)=(1,11)$. The generator network consisted of three hidden layers with ReLU–ReLU–Tanh activations and a Sigmoid output layer, with layer sizes $500$, $1000$, $500$, and $4n$ with input dimension $=10$. The discriminators used ReLU activations with a linear output. Training was performed using the Adam optimizer with learning rate $2\times10^{-4}$. The training performance is shown in Fig.~\ref{trainingloss}, where the generator loss decreases steadily over training iterations, reaching convergence by epoch~5. In the post-training evaluation, stealthiness values lie below the threshold ($\tau_s = 0.04$), while effectiveness values lie above the threshold ($\tau_e = 0.64$), based on a 25\% subset of generated samples, compared to pre-training performance, thereby validating the effectiveness of the attack generation framework.

\begin{figure}[htpb]
    \centering
    \includegraphics[scale=0.65]{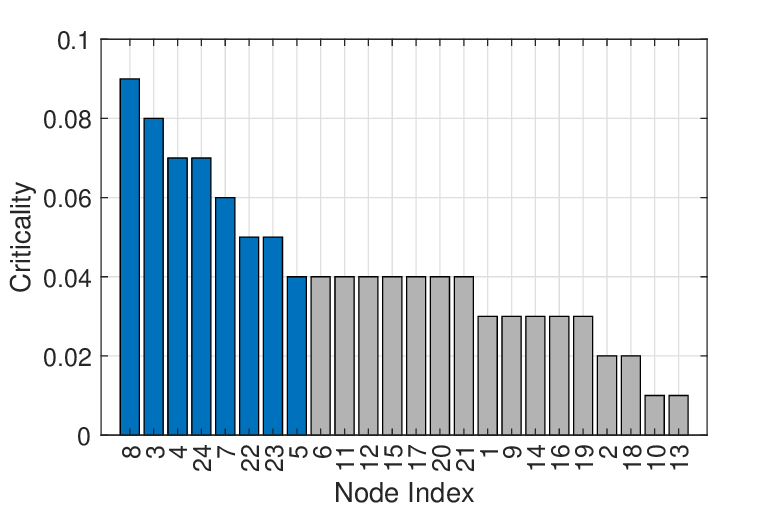}
    \caption{Top-8 ranked grid sensor node criticality scores.}
    \label{node_critic}
\end{figure}

\begin{figure}[htpb]
    \centering
    \includegraphics[scale=0.7]{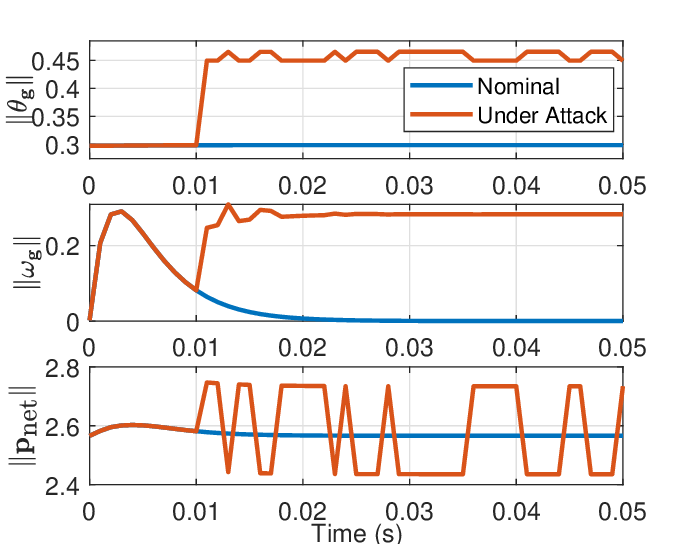}
    \caption{Grid angle, frequency, and net power responses under random attack injections targeting critical nodes.}
    \label{meas_impac_nom_node}
\end{figure}

\begin{figure}[htpb]
    \centering
    \includegraphics[scale=0.7]{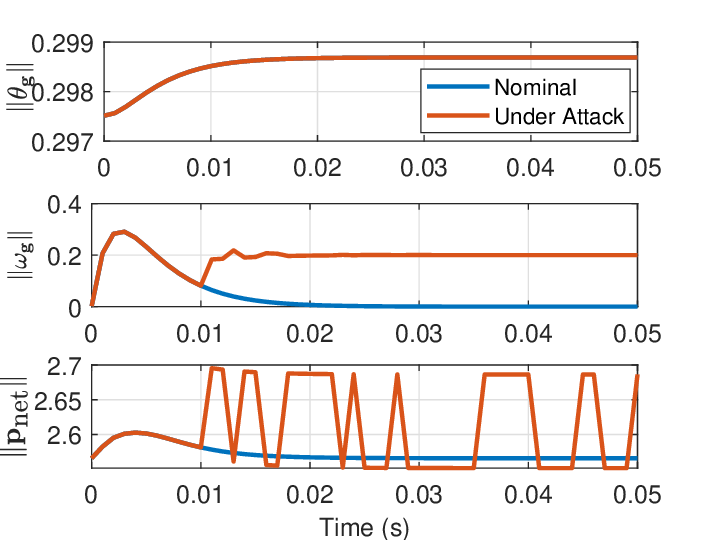}
    \caption{Grid angle, frequency, and net power responses under random attack injections targeting randomly selected nodes.}
    \label{meas_impact_rand_node}
\end{figure}

The node criticality assessment was performed using \( M = \left(\frac{n}{2}\right)! \) sampled permutations of the attackable node set. 
The results of the node criticality assessment for the power grid model are presented in Fig. \ref{node_critic}, which ranks the grid sensor nodes based on their computed criticality scores. A clear disparity is observed among nodes, with nodes 8, 3, 4, 24, 7, 22, 23, and 5 exhibiting the highest importance, indicating their potential influence on overall grid stability. To validate these findings, targeted attacks were conducted on both the identified critical nodes and randomly selected nodes. The resulting system responses measured in terms of grid angle, frequency, and net power deviations are shown in Fig. \ref{meas_impac_nom_node} and Fig. \ref{meas_impact_rand_node}, respectively. As shown, system responses exhibited noticeably larger deviations from nominal behavior when attacks targeted critical nodes, whereas responses to random attacks were comparatively less severe.

To quantify these differences, the energies associated with response deviations were computed, as summarized in Table \ref{Metrics}. The results indicated that the criticality-based node selection strategy consistently resulted in greater system degradation than random selection across all measured signals. In particular, generator frequency and network power deviations increased by approximately 50\%, while the overall system output $\|\mathbf{y}\|$ showed a smaller increase of about 5\%. For generator angle deviations, random selection yielded no impact, whereas critical selection produced a nonzero deviation, underscoring its effectiveness. Finally, Table \ref{Metrics2} lists the top five most critical buses (3, 6, 2, 1, 8), which were obtained by computing bus-level criticality as a linear combination of associated measurement-node values and ranking the resulting scores in descending order.
\begin{table}[htpb]
\caption{Power measurements degradation under criticality-based and random node selection strategies.}
\centering
\begin{tabular}{c|c|c}
\hline
Signal & Critical Selection & Random Selection  \\ 
\hline
$\| \boldsymbol{\theta}_g \|$ &0.03   &0.00   \\
$\| \boldsymbol{\omega}_g  \|$ &0.06  &0.04  \\
$\| \mathbf{p}_{\text{net}} \|$ &0.03  & 0.02 \\
$\| \mathbf{y} \|$ &0.60  &0.57  \\
\hline
\hline
\end{tabular}
\label{Metrics}
\end{table}
\begin{table}[htpb]
\caption{Top-5 most critical buses identified from node criticality assessment.}
\centering
\begin{tabular}{c|c|c}
\hline 
Nodes & Bus  & Critical values \\ 
\hline
 3, 8, 13 & 3& 0.18   \\ 
 4, 9, 16&  6&  0.13 \\ 
2, 7, 12 &  2&   0.12\\ 
1, 6, 11 &  1&   0.11\\ 
5, 10, 18 &  8&   0.07\\ 
\hline
\hline
\end{tabular}
\label{Metrics2}
\end{table}

\subsection{Scalability Considerations}
A practical limitation of the proposed node criticality framework is the combinatorial growth of coalition permutations as the number of admissible measurement nodes increases. For $n$ nodes, exact evaluation requires $n!$ permutations, which rapidly becomes computationally intractable for large-scale power systems. To mitigate this challenge, the framework employs reduced permutation sampling, where only a subset $M \ll n!$ of permutations is evaluated to approximate node marginal contributions. This significantly lowers computational burden while preserving reliable criticality estimates. Additional efficiency may be achieved through heuristic node pre-screening, coalition size  constraints, or parallelized implementation. These strategies provide a practical trade-off between computational scalability and ranking accuracy, enabling application of the proposed framework to larger cyber--physical grid networks.

\section{Conclusion}
This paper presented an adversarially informed framework for identifying critical measurement nodes in power grids by jointly integrating network dynamics, structured attack generation, and resilience-aware state estimation. Unlike conventional topology or flow-based ranking methods, the proposed approach explicitly accounts for worst-case measurement corruption under stealth and effectiveness constraints, enabling a more operationally meaningful notion of node criticality. The results demonstrate that nodes identified under the proposed framework induce higher degradation in system performance compared to random selections, confirming the relevance of adversarial considerations in grid vulnerability assessment.

\section*{Code Availability}
The source code used to generate the simulation results presented in this paper
is made publicly available at:
https://github.com/resilient-autonomous-systems-lab.

\section*{Acknowledgment}
This work was supported by the U.S. Department of Energy
under Award Number DE-CR0000028. The authors
acknowledge that the views expressed do not necessarily
reflect those of the United States Government.

\bibliographystyle{IEEEtran}
\bibliography{myreferences}

\end{document}